\documentclass[twocolumn,pre,aps,showpacs,amsmath,amssymb,superscriptaddress]{revtex4}

\usepackage{hyperref}
\usepackage{amsmath,amssymb}
\usepackage{amsfonts,amsthm}
\usepackage{graphics}
\usepackage{graphicx}
\usepackage{dcolumn}
\usepackage{color}
\usepackage{bm}
\usepackage[dvipsnames]{xcolor}
\usepackage[normalem]{ulem}
\usepackage[bf]{subfigure}

\begin{document}

\title{A symbolic information approach applied to human intracranial data to characterize and distinguish different congnitive processes
}

\author{Ícaro Rodolfo Soares Coelho Da Paz}
\affiliation{Instituto de F\'{\i}sica, Universidade Federal de Alagoas, Macei\'{o}, Alagoas 57072-970 Brazil.}

\author{Pedro F. A. Silva} 
\affiliation{Instituto de F\'{\i}sica, Universidade Federal de Alagoas, Macei\'{o}, Alagoas 57072-970 Brazil.}

\author{Helena Bordini de Lucas}
\affiliation{Instituto de F\'{\i}sica, Universidade Federal de Alagoas, Macei\'{o}, Alagoas 57072-970 Brazil.}

\author{Sérgio H. A. Lira}
\affiliation{Instituto de F\'{\i}sica, Universidade Federal de Alagoas, Macei\'{o}, Alagoas 57072-970 Brazil.}

\author{Osvaldo A. Rosso}
%\thanks{oarosso@fis.ufal.br}
\affiliation{Instituto de F\'{\i}sica, Universidade Federal de Alagoas, Macei\'{o}, Alagoas 57072-970 Brazil.}

\author{Fernanda Selingardi Matias}
\thanks{fernanda@fis.ufal.br}
\affiliation{Instituto de F\'{\i}sica, Universidade Federal de Alagoas, Macei\'{o}, Alagoas 57072-970 Brazil.}

\begin{abstract}

How the human brain processes information during different cognitive tasks is one of the greatest
questions in contemporary neuroscience. Understanding the statistical properties of brain signals during specific activities is one promising way to address this question. Here we analyze freely available data from
implanted electrocorticography (ECoG) in five human subjects during two different cognitive tasks
in the light of information theory
quantifiers ideas. We employ a symbolic information approach to determine the probability distribution function associated with the time series from different cortical areas. Then we utilize these
probabilities to calculate the associated Shannon entropy and a statistical complexity measure based on the disequilibrium between the actual time series and one with a uniform probability distribution function. We show that an Euclidian distance in the complexity-entropy plane and an asymmetry index for complexity are useful for comparing the two conditions.
We show that our method can distinguish visual
search epochs from blank screen intervals in different electrodes and patients. By using a multi-scale approach
and embedding time delays to downsample the data, we find important time scales in which the relevant information is being processed.
We also determine cortical regions and time intervals along the 2-second-long trials that present more pronounced differences between the two cognitive tasks.
Finally, we show that the method is useful to
distinguish cognitive processes using brain activity on a trial-by-trial basis.

\end{abstract}

% \pacs{87.18.Sn, 87.19.ll, 87.19.lm}

\maketitle

\section{Introduction}
\label{Sec-Introduction}

%%%%%%%%%%%%%%%%%%%%%%%%%%%%%%%%%%%%%%%%%%%%
Understanding the statistical properties of the human brain activity during different cognitive processes is a great step toward comprehending how the brain processes information. Characterizing these properties in different cortical regions, and time scales can contribute to this issue.
Electrophysiological data from intracranial electrodes in the human brain
are promising to address these questions since it
provides millisecond temporal resolution, clear signals from specific brain areas, and a good signal-to-noise rate~\cite{parvizi2018promises}.
However, the overall access to intracranial data
is still rare and remains somewhat exclusive to the experimentalists who have produced it.

Here, we analyze freely available data from implanted	
electrocorticographic (ECoG) measurements
of brain surface potentials in five human subjects during visual cognitive tasks~\cite{miller2010dynamic}.
It has been previously reported for this data that 
a comparison of
visual search trials with interspersed blank screen intervals presents changes in the raw potential, in the amplitude of rhythmic activity, and in the decoupled broadband spectral amplitude~\cite{miller2010dynamic}.
In the present work, we extend this comparison between active visual search tasks and waiting windows by using statistical tools that can discriminate which brain areas are more engaged in one of the trial types during specific time intervals.  

To characterize different features of time series we employ two time causal quantifiers based on Information Theory: Shannon entropy (see Eq.~\ref{eq:Shannon-entropy})\cite{Shannon}, and the corresponding Martín-Platino-Rosso (MPR) statistical complexity (see Eq.~\ref{eq:Complexity}), based on the disequilibrium between the actual time series and one with a uniform probability distribution function
\cite{lopez1995statistical,Lamberti04,Rosso07,Zunino12,Xiong2020}. Using the MPR definition of complexity, both extremes of order and disorder present low complexity. For example, a constant time series or a very noisy time series would present low complexity. In the same fashion, a perfect crystal or a random distribution of atoms are not complex systems. We assign each time series under study a
position in a two-dimensional space spanned by the entropy
and the statistical complexity measure: the complexity-entropy plane $C\times H$.
These quantifiers are evaluated using the Bandt-Pompe symbolization methodology~\cite{Bandt02}, which includes naturally the time causal ordering provided by the time-series data in the corresponding associate probability distribution function (PDF). 

This approach was originally
introduced to distinguish chaotic from stochastic systems in
time series analysis~\cite{Rosso07,Zunino12}. Recently, it has been successfully applied to study brain signals: to estimate time differences during phase synchronization~\cite{Montani15}, to show that complexity is maximized close to criticality in cortical states~\cite{Lotfi21,capek2023parabolic},
to distinguish cortical states using EEG data~\cite{baravalle2018rhythmic,mateos2021using}, to characterize neurological diseases using MEG~\cite{echegoyen2020permutation},
as well as to study neuronal activity~\cite{Montani2015neuronas,Montani15,deLuise2021network}.
Furthermore, it has been applied to monkey LFP to estimate response-related differences between Go and No-Go trials~\cite{deLucas21}.
As far as we know it has not been applied to intracranial human data before.

Here we show that our method can distinguish 
visual search epochs from blank screen intervals in specific brain regions, for different time intervals, and at relevant time scales.
The experimental data corresponding to ECoG, are presented
In Sec.~\ref{Sec:Experimental-data}. The Bandt-Pompe symbolization method is explained in Sec.~\ref{Sec:BP-method}.
In Sec.~\ref{Sec:Information-quantifiers} we introduce the Information Theory quantifiers employed in our analysis.
In Sec.~\ref{Sec:Results}, we report our results for all 67 electrodes, focusing on the ones with pronounced differences between trial types from different patients. At the end of this section,
we also show that the method can distinguish two cognitive tasks on a trial-by-trial basis.
Finally, concluding remarks, perspectives, and a brief discussion of the significance of our findings for neuroscience are presented in Sec.~\ref{Sec:Conclusions}.

\section{Experimental data}
\label{Sec:Experimental-data}

%%%%%%%%%%%%%%%%%%%%%%%%%%%%%%%%%%%%%%%
\begin{figure}[t]
% \begin{minipage}{8cm}
%  \begin{flushleft}(a)%
%\end{flushleft}%
%\includegraphics[width=0.96\columnwidth,clip]{figura_task.png}
%\end{minipage}
 \begin{minipage}{8cm}
  \begin{flushleft}(a)%
\end{flushleft}%
\includegraphics[width=0.96\columnwidth,clip]{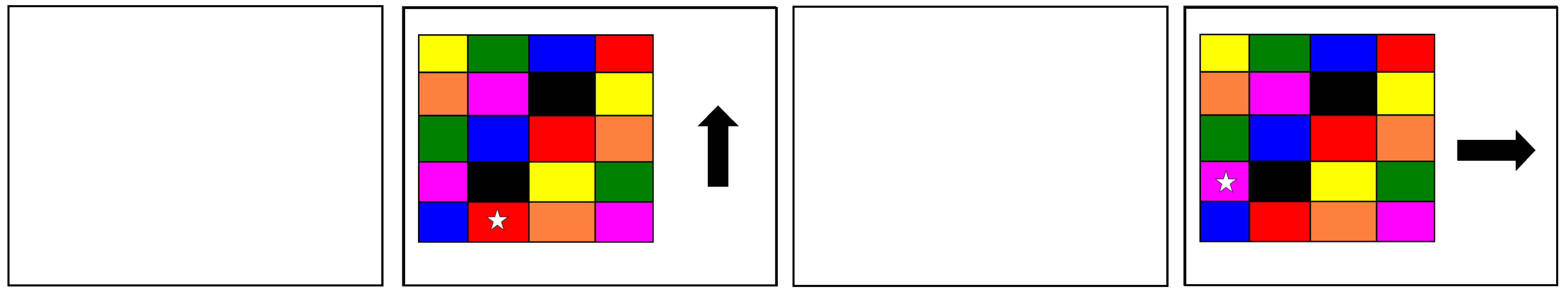}
\end{minipage}
\begin{minipage}{8cm}
\begin{flushleft}(b)%
\end{flushleft}%
\includegraphics[width=0.98\columnwidth,clip]{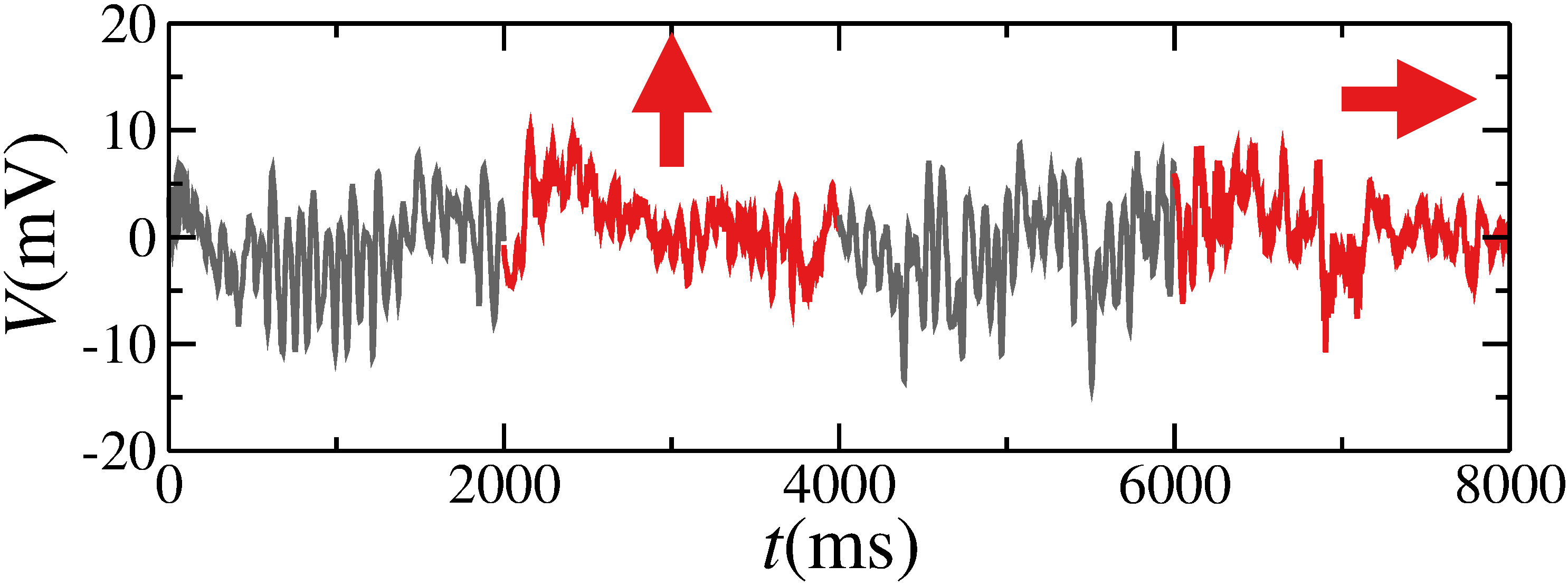}
\end{minipage}
\caption{
Experimental paradigm. (a) Illustration of the screen during the time course of the experiment: 2000~ms of waiting periods observing a white screen (called \textit{blank}) alternated with 2000~ms of an active visual search task (called \textit{search}). (b) Time series of the recorded activity of an exemplar electrode (JT3) at an occipital area during the two \textit{blank} screens and the two \textit{search} periods shown in (a).}
\label{fig:trials}
\end{figure}

%%%%%%%%%%%%%%%%%%%%%%%%%%%%%%%%%%%%%%%%%%%%%%%%%

We	have analyzed freely available data from 67 implanted electrocorticographic electrodes in five human subjects during a visual cognitive paradigm~\cite {miller2010dynamic}. %pedem para colocar exatamente assim 
Ethics statement: All patients participated in a purely voluntary manner, after providing informed written consent, under experimental protocols approved by the Institutional Review Board of the University of Washington ($\#$12193). All patient data was anonymized according to IRB protocol, in accordance with HIPAA mandate. These data originally appeared in the manuscript “Dynamic modulation of local population activity by rhythm phase in human occipital cortex during a visual search task” published in Frontiers in Human Neuroscience in 2010~\cite{miller2010dynamic}.

Sub-dural grids and strips composed of platinum electrodes were strategically positioned over the frontal, parietal, temporal, and occipital cortex regions to facilitate prolonged clinical observation and accurate localization of seizure foci. 
 The electrodes, featuring a 4~mm diameter and $+$ 1~cm inter-electrode distance, were utilized for data collection. The potentials were sampled at a rate of 1000~Hz (one point every 1~ms) in relation to a scalp reference and ground. These signals underwent an instrument-imposed band-pass filter, ranging from 0.15 to 200~Hz.

The experimental procedure involved two different visual cognitive tasks in which subjects engaged with stimuli displayed on an LCD monitor positioned 1 meter away. The active \textit{search} task 
comprised 120 trials, each consisting of 2000~ms visual search stimuli (referred to hereafter as \textit{search} trials) interleaved with 2000~ms inter-stimulus intervals (ISIs), during which the screen remained blank (a simple waiting task referred to hereafter as \textit{blank} trials). See Fig.~\ref{fig:trials}(a) for an example of the observed sequence of intercalated trial types in the screen.
Each visual search stimulus comprised three elements: (1) a 5-row by 4-column array of colored boxes, each measuring 1~cm by 1~cm, (2) a white star positioned at the center of one of these boxes, and (3) a black arrow (measuring 2~cm by 1~cm) situated 1.5~cm to the right of the right-most box in the middle row. The placement of the star and arrow within the colored boxes was randomized, with the arrow pointing in one of four cardinal directions: ``right", ``left", ``up", or ``down". Participants were tasked with identifying the color of the box adjacent to the star in the direction indicated by the arrow. For instance, in Fig.~\ref{fig:trials}(a), the two examples of visual tasks should have been correctly answered as ``black" and ``black". They repeat the task for 120 trials (30 trials for each direction) ~\cite{miller2010dynamic}. 

As a first step to our data analysis, for each electrode, we have separately concatenated the time series of the 120 ISIs (\textit{blank} trials)
and the 120 stimuli (\textit{search} trials). In Fig.~\ref{fig:trials}(b) we show the electric activity of an exemplar electrode at the occipital cortex for two trials of the blank screen (before concatenation) alternated by two trials of visual search task (one for an ``up" arrow and other for a ``right" arrow). 
Therefore, for each channel and each trial type, we determine the information theory indexes associated with its time series as described below.

\section{The Bandt-Pompe symbolization method}
\label{Sec:BP-method}

%%%%%%%%%%%%%%%%%%%%%%%%%%%%%%%%%%%%%%%%%%%%%%%%%
\begin{figure}[t]
 \begin{minipage}{8cm}
  \begin{flushleft}(a)%
\end{flushleft}%
\includegraphics[width=0.96\columnwidth,clip]{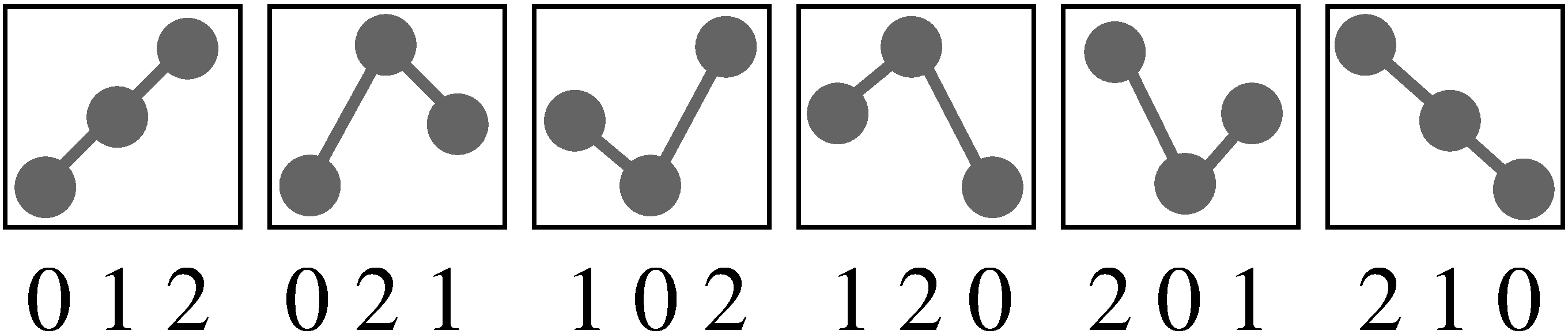}
\end{minipage}
\begin{minipage}{8cm}
\begin{flushleft}%
\end{flushleft}%
\includegraphics[width=0.98\columnwidth,clip]{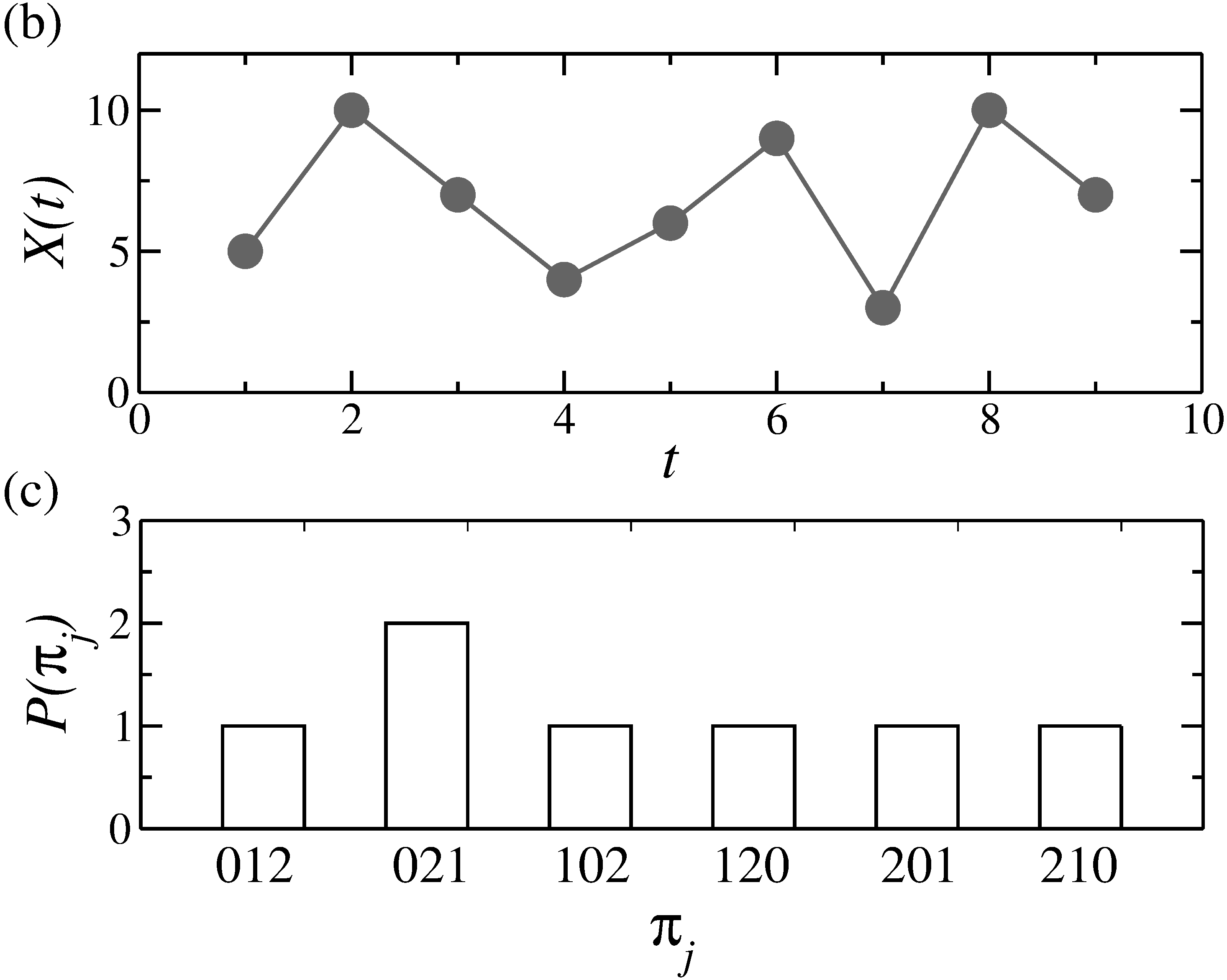}
\end{minipage}
\caption{\label{fig:DSAS} 
Characterizing the symbolic representation of time series.
(a) The six possible symbols associated with permutations $\pi_j$ for ordinal patterns of length $D=3$.
(b) Example of a very simple time series $X(t)=\{5,10,7,4,6,9,3,10,7\}$ and (c) its own non-normalized probability density function (PDF).
}
\label{fig:symbol}
\end{figure}

%%%%%%%%%%%%%%%%%%%%%%%%%%%

In order to calculate any information theory
quantifier, one should obtain a probability distribution function (PDF) from a time series
$X(t)=(x_{1},x_{2},x_{3}, \cdots)$ representing the evolution dynamics of the system
under study, where $x_j$ is the studied variable evaluated at time $t_j$, and the colection of $t_j$ are usually equally spaced. 
There is no unique answer for the best procedure to associate a time series with a PDF, and in fact, different proposals can be found.
Here, we use a symbolic representation of a time series introduced by Bandt \& Pompe (BP)~\cite{Bandt02} for evaluating the PDF.  This symbolization techinique consists of extracting the ordinal patterns of
length $D$, associated to each time $t$ of our time series, generated by $\textbf{s}(t)=(x_{t},x_{t+1}, \cdots ,x_{t+D-1},x_{t+D})$. This corresponds to indexing each $t$ to the $D$-dimensional vector $\textbf{s}(t)$.
The greater the value of $D$, the more information is incorporated into the vectors. 

We should identify and count the number of occurrences of
all $D!$ symbols $\pi_j$ of length $D$ (with $j=1,2,...,D!$, see the six $\pi_j$ for $D=3$ in Fig.~\ref{fig:symbol}(a)).
The specific $j-th$ ordinal pattern associated to $\textbf{s}(t)$ is the permutation $\pi_j=(r_{0},r_{1},...,r_{D-1})_j$ of $(0,1,...,D-1)$ which guarantees that $x_{t+r_{0}} \leqslant x_{t+r_{1}} 
  \leqslant  \cdots  \leqslant x_{t+r_{(D-2)}} \leqslant x_{t+r_{(D-1)}} $. 
In order to get a unique result, we set $r_i < r_{i+1}$ if $x_{t+r_i} = x_{t+r_{i+1}}$. We follow the chronological index permutation
mapping for $\pi_j$~\cite{Trevesaro2018}.
In other words, each permutation $\pi_j$ is one of our possible symbols and we have $D$! different symbols .
Therefore, the pertinent symbolic data is created by the following rules: (i) grouping the $D$ consecutive values of the time series points in the vector $\textbf{s}(t)$, (ii) indexing a symbol $\pi_j$ to the vector $\textbf{s}(t)$ by reordering the embedded data in ascending order using the permutation $\pi_j$. Therefore, for each $x_t$ (with $t=1,2, \dots ,M-(D-1)$), we can associate a symbol $\pi_j$. 

Afterward, it is possible to quantify the diversity of the ordering symbols (patterns) derived from a scalar time series by counting how many times each one of the $D$! different permutations $\pi_j$ have been found in the data-set. Then, to calculate the PDF (for a specific $D$), we find
$P\equiv \{p_j;j=1,2,...,D! \}$, where $p_j$ is the probability to find the $j$-th symbol $\pi_j$ in our time series. This procedure is essential to a phase-space reconstruction with embedding dimension (pattern length) $D$. For practical purposes, BP suggested to use $ 3\leqslant D \leqslant 7 $.

To have an example, choosing $D=3$, all the 6 possible symbols associated with the permutations $\pi_j$ are presented in Fig.~\ref{fig:symbol}(a). Considering the time series $X(t)=\{5,10,7,4,6,9,3,10,7\}$ as an example (see Fig.~\ref{fig:symbol}(b)), the first vector is $\textbf{s}(t=1)=(5,10,7)$, corresponding to the permutation $\pi_2=(0,2,1)$; the second vector is $\textbf{s}(t=2)=(10,7,4)$, corresponding to to the permutation $\pi_6=(2,1,0)$. Similarly, one can find the other five vectors $\textbf{s}(t)$ and its respective $\pi_j$. The correspondent non-normalized PDF is shown in Fig.~\ref{fig:symbol}(c).

We can also repeat the analysis including a time embedding (also called a time delay) $\tau$ to evaluate the PDF in different time scales. In order to do that, we skip every $\tau-1$ points of our time series $X(t)$ in order to find and count the symbols. In the previous example we use $\tau=1$ and consider every point in $X(t)$. For $\tau=2$ we would skip every other point in such a way that the first vector is $\textbf{s}(t=1)=(5,7,6)$, corresponding to the permutation $\pi_2=(0,2,1)$; the second vector is $\textbf{s}(t=2)=(10,4,9)$, corresponding to the permutation $\pi_5=(2,0,1)$; the third vector is $\textbf{s}(t=3)=(7,6,3)$, corresponding to the permutation $\pi_6=(2,1,0)$. Therefore, to each time series $X(t)$ we can associate many PDFs, each one for a different value of the time delay $\tau$. Unless otherwise stated here we evaluate the PDF for $1\leq \tau \leq 50$. Since the sampling rate of our data is $1000$~Hz, $\tau$ can be represented in miliseconds.

It is noteworthy that the symbol sequences emerge from the time series without the need for model-based assumptions. Although this approach may result in some loss of fine-grained details pertaining to the original series' amplitude, it effectively captures the temporal structure of the time series, providing insights into the system's temporal correlation [31,32].
It is important to emphasize that the BP methodology only requires a weak stationarity assumption:  for $k \leq D$, the probability for $x_t \leq x_t+k$ should not depend on $t$.

\section{Information Theory quantifiers: permutation entropy and statistical complexity}
\label{Sec:Information-quantifiers}

After associate a probability distribution function
(PDF) to the time series, we can evaluate the corresponding Information Theory quantifiers. The first Information Theory index that we introduce is the Shannon's logarithmic information measure \cite{Shannon49}, defined by:
\begin{equation}
\label{eq:Shannon-entropy}
S[P] = -\sum_{j=1}^{N} p_j \ln(p_j) \ .
\end{equation}
This functional is equal to zero when we are able to correctly predict the outcome every time. For example, for linearly increasing time series all probabilities are zero but one which is equal to 1. The corresponding PDF will be $P_0 = \{ p_k =1 ~{\mathrm {and}} ~p_j = 0, \forall j\neq k, j = 1, \dots, N-1 \}$, then $S[P_0]=0$. 
By contrast, the entropy is maximized for the uniform distribution $P_e = \{p_j=1/N, \forall j=1,2, \dots,N\}$, being $S_{max}= S[P_e]= \ln( N )$.
We define the normalized Shannon entropy by
\begin{equation}
\label{eq:Normalized-Shannon-entropy}
H[P] = \frac{S[P]}{S_{max}} = \frac{S[P]}{\ln(N)} \ ,
\end{equation}
and $0 \leq H[P]\leq 1$, which give a measure of the information content of the corresponding PDF ($P$).

The second Information Theory based quantifier that we introduced is the Statistical Complexity, defined by functional product form~\cite{lopez1995statistical,Rosso07}
\begin{equation}
\label{eq:Complexity}
C[P, P_e] = H[P] \cdot  Q_J[P,P_e] \ .
\end{equation}
Where, $H[P]$ is the normalized Shannon entropy (see Eq.~(\ref{eq:Normalized-Shannon-entropy})) and $Q_J[P,P_e]$ represent the disequilibrium, which is defined in terms of the Jensen–Shannon divergence ~\cite{Grosse02} as:
\begin{equation}
\label{eq:Q-disequilibrium}
Q_J[P,P_e] = Q_0 J[P,P_e] \ ,
\end{equation}
where
\begin{equation}
\label{eq:Jensen-Shannon-Divergence}
J[P,P_e] = S\left[\frac{(P+P_e)}{2}\right] - \frac{S[P]}{2} - \frac{S[P_e]}{2}  \ ,
\end{equation}
and $Q_0$ is a normalization constant ($0\leq Q_J \leq 1$), equal to the inverse of the maximum possible value of $J[P,P_e]$, that is $Q_0 = 1 / J[P_0,P_e]$.
In this way, also the statistical complexity is a normalized quantity, $0\leq C[P,P_e] \leq 1$.
It is interesting to note, that the statistical complexity give additional information in relation to the entropy, due to its dependence on two PDFs. Moreover, it can be shown that for a given value of the normalized entropy $H$, the corresponding complexity varies in a range of values given by $C_{\mathrm{min}}$ and $C_{\mathrm{max}}$, and these values depend only on the dimension of the PDF considered and the functional form chosen for the entropy ~\cite{Martin06}.

Therefore, we can define a distance between entropy and complexity for two conditions $i$ and $j$ as the simple Euclidean distance:
\begin{equation}
\label{eq:distance}
    d_{i,j}=\sqrt{ (H_{i} - H_{j})^2 + (C_{i}-C_{j})^2}.
\end{equation}
We can also calculate a normalized asymmetry index for complexity as~\cite{deLucas21}:
\begin{equation}
\label{eq:asymmetric-C}
    A(C_{search},C_{blank})=\frac{ C_{search} - C_{blank}} {C_{search} + C_{blank}}  \ .
\end{equation}
Both $d_{blank,search}$ and $A(C_{search},C_{blank})$ are independently calculated for each $\tau$.

Applying this methodology to the electrical activity from each channel, we calculate entropy $H_i$, and complexity $C_i$ for each trial type ($i=blank,task$) as a function of the time embedding $\tau$ using time series with 240000 time points (120 trials x 2000 miliseconds).
Then we apply a sliding window in each 2000~ms trial, dividing it into smaller windows of 100~ms and 200~ms (starting every 10~ms) and concatenating the correspondent i-th window of each trial of the same type. For this analysis, our smallest time series has 3000 time points (30 trials x 100 milliseconds). Results are consistent and qualitatively similar when comparing 100~ms and 200~ms windows. Therefore, we show the results for 200~ms windows.
In figures related to the sliding window, we associate each window with the time of the beginning of the window. This means, for example, that the entropy at $t=500$~ms is in fact the entropy calculated during the interval from 500 to 700~ms.

Finally, we analyze the data on a trial-by-trial basis. Using the 2000 points to obtain 120 values of $H_{blank}$, $C_{blank}$, $H_{search}$, and $C_{search}$ (one for each trial of the spcific type). We have also applied the sliding window to each trial and calculated $H$ and $C$ using only 200 points for each 200~ms long window. Results are qualitative similar among different values of the length of the ordinal pattern $D$ which determines the number of possible symbols $D!$. Unless otherwise stated, we show the results for $D=6$.

%%%%%%%%%%%%%%%%%%%%%%%%%%%%%%%%%%%%%%%%%%%%%%%%%

\begin{figure}[t]%
% \begin{minipage}%{8cm}
  \begin{flushleft}%
\end{flushleft}%
\includegraphics[width=0.99\linewidth,clip]{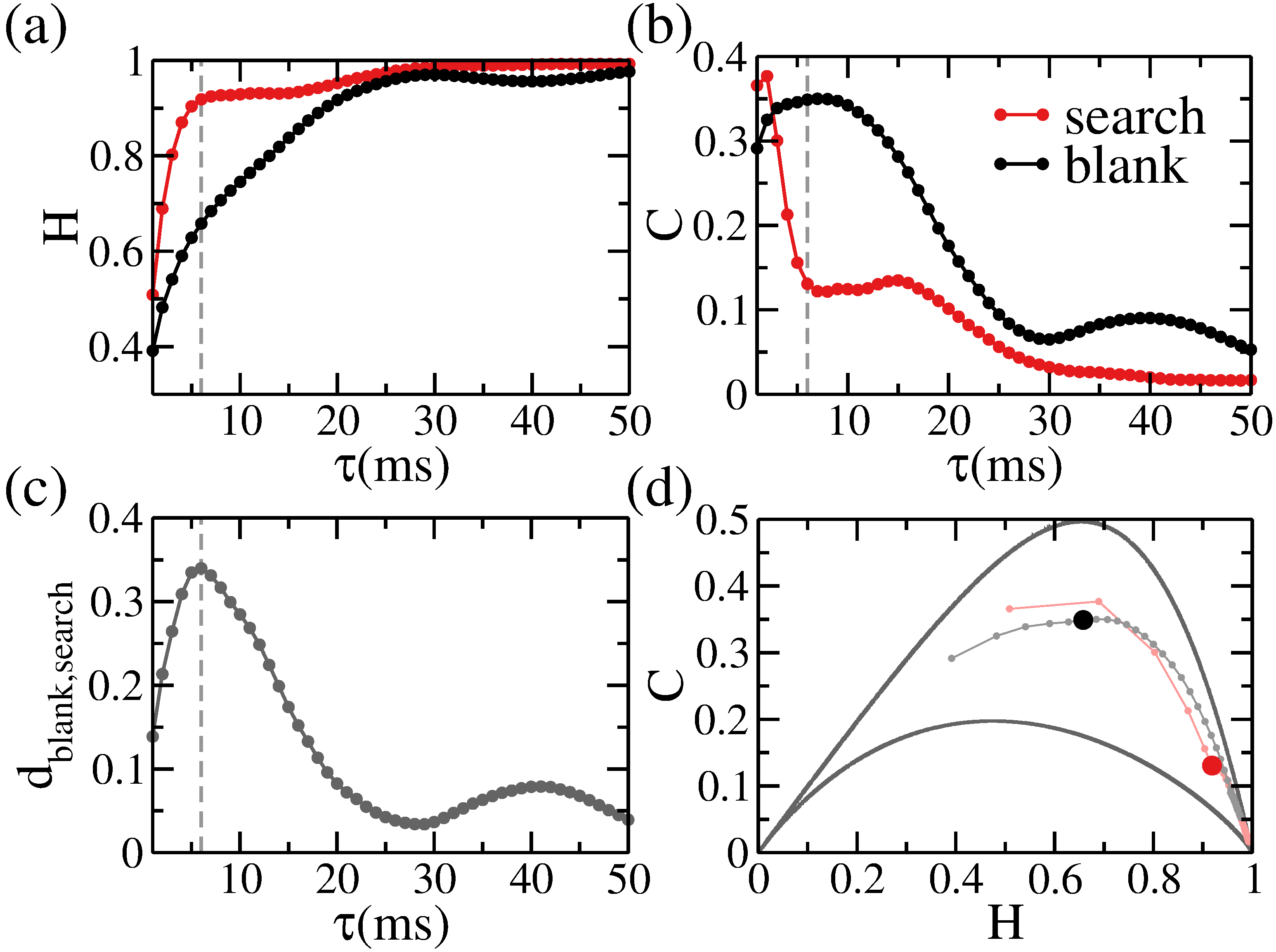}
%\end{minipage}
\caption{\label{fig:HCjt3} Information theory quantifiers to characterize different cognitive tasks: waiting window during a \textit{blank} screen trial (black) and visual \textit{search} task (red).
(a) Entropy $H$ and
(b) Statistical complexity $C$ as a function of the embedding time delay $\tau$. 
(c) Euclidean distance $d_{blank,search}$ in the $C\times H$ plane between the two tasks (see Eq.\ref{eq:distance})
(d) Complexity-entropy plane highlighting $\tau=6$~ms which maximizes the distance for this exemplar electrode. The indexes were calculated using the time series which is partially shown in Fig.~\ref{fig:trials}(b).
}
\end{figure}

%%%%%%%%%%%%%%%%%%%%%%%%%%%%%%%%%%%%%%%%%%%%%%%%%

\section{Results}
\label{Sec:Results}

\subsection{Applying information theory quantifiers to characterize and distinguish trial types}

We employ information theory quantifiers $H$ and $C$ to study brain signals and characterize cortical information processes during an experimental paradigm related to two visual cognitive tasks.  
We show that our method is able to distinguish the two conditions: visual \textit{search} epochs related to an active task, from \textit{blank} screen intervals related to waiting periods. We show that some cortical regions and specific time scales exhibit more pronounced differences between the two trial types:\textit{blank} and \textit{search}. We have separately analyzed the time series of all \textit{blank} trials, as well as of all \textit{search} trials for the 67 electrodes of five patients.

An illustrative example of the time series with the electrical activity of a few trials for an exemplar electrode (called JT3) is shown in Fig.~\ref{fig:trials}(b). First, we calculate entropy $H_{blank}$ and complexity $C_{blank}$ as a function of the embedding time $\tau$ for all the 120 \textit{blank} trials concatenated along the total 2 s duration of each trial
(see black curves in Fig.~\ref{fig:HCjt3}(a) and (b) and Sec.~\ref{Sec:Experimental-data} for more details). Then we calculate entropy $H_{search}$ and complexity $C_{search}$ in a similar way for all the 120 \textit{search} trials concatenated (see red curves in Fig.~\ref{fig:HCjt3}(a) and (b)). 

Regarding the effect of using the embedding time $\tau$ to explore multiple time scales from 1~ms up to 50~ms, three following results should be emphasized.
i) The indexes $H$ and $C$ present nontrivial dependence with the embedding time $\tau$ which indicates that the information process occurring in this cortical region is different for different time scales.
ii) Some time scales seem to work better for distinguishing \textit{search} trials from \textit{blank} trials. Especially for $3$~ms~$<\tau<12$~ms there is a clear separation of the indexes regarding the trial type. 
iii) Larger values of $\tau$ ($\tau>30$~ms) present larger $H$ and smaller $C$,
for both \textit{blank} and \textit{search} trial types,
which characterizes noisier time series.

Furthermore, for each $\tau$ we calculate the Euclidean distance $d_{blank,search}$ in the complexity-entropy plane between the two trial types (see Fig.~\ref{fig:HCjt3}(c) and Eq.~\ref{eq:distance}).
For the exemplar electrode JT3, the distance is maximized arround $\tau=6$~ms. 
We can use the 2D multi-scale complexity-entropy plane ($C\times H$) as an extra way to visualize our results (see Fig.~\ref{fig:HCjt3}(d)). In this plane, we can better characterize each cognitive task as well as visualize how the separation between different time series changes as $\tau$ is varied. For instance, in Fig.~\ref{fig:HCjt3}(d), the highlighted dots depict the positions in the $C\times H$ plane for \textit{blank} and \textit{search} for a fixed $\tau=6$~ms, and the distance between them can be readly visualized. As we increase $\tau$, the points in the $C\times H$ plane move along the grey and red lines, where both move in the direction of the lower right corner related to noisier series.

%%%%%%%%%%%%%%%%%%%%%%%%%%%%%%%%%%%%%%%%%%%%%%%%%%%%%%%%%
\begin{figure}[t]%
 %\begin{minipage}{8cm}
  \begin{flushleft}%
\end{flushleft}%
\includegraphics[width=0.999\columnwidth,clip]{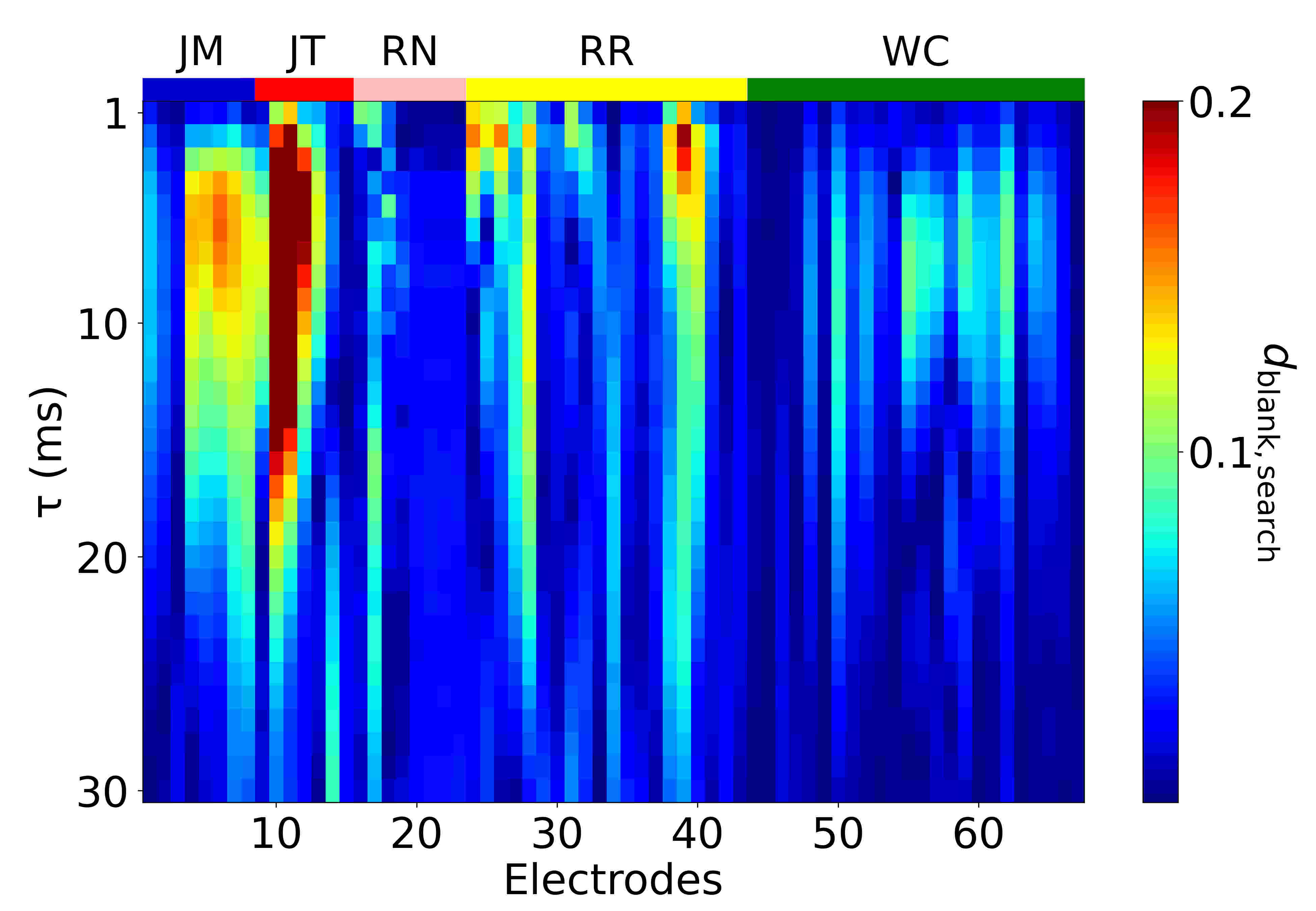}
%\end{minipage}
\caption{\label{fig:heatmapallchannels} 
Identifying which channels and time scales can better distinguish the two cognitive tasks: \textit{blank} screen and visual \textit{search} task.
Heatmap representing the distance $d_{blank,search}$ in the complexity-entropy plane between two trial types, for all 67 electrodes on the y-axis and time delays $\tau$ up to $30$~ms in the x-axis.   
}
\end{figure}

%%%%%%%%%%%%%%%%%%%%%%%%%%%%%%%%%%%%%%%%%%%%%%%%%

\begin{figure*}[t]%
 %\begin{minipage}{8cm}
 % \begin{flushleft}%
%\end{flushleft}%
\includegraphics[width=0.999\linewidth,clip]{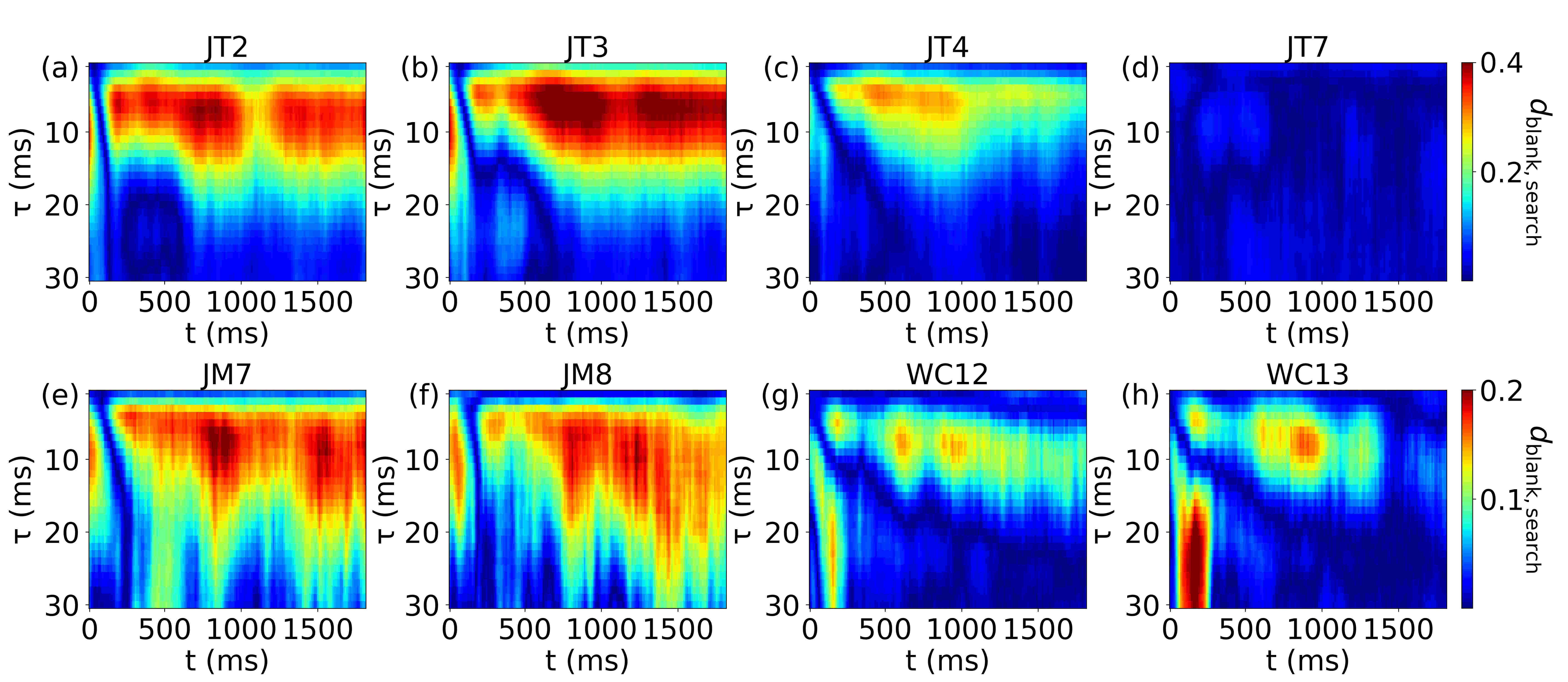}
%\end{minipage}
% \begin{minipage}{8cm}
%  \begin{flushleft}%
%\end{flushleft}%
%\includegraphics[width=0.98\columnwidth,clip]{JT - Canal 3 - 500-700.png}
%\end{minipage}
\caption{\label{fig:Heatmap8channels} 
Differences between the two cognitive tasks along different time windows of the 2-second-long trial and multiple time-scales. Heatmap representing the distance $d_{blank,search}$ in the $C\times H$ plane between \textit{blank} and \textit{search} trials for 8 illustrative channels from different patients. On the vertical axis, we show the time embedding $\tau$ and on the horizontal axis, we show the time course of the 2-second-long trials.
}
\end{figure*}

%%%%%%%%%%%%%%%%%%%%%%%%%%%%%%%%%%%%%%%%%%%%%%%%%
\begin{figure*}%
\includegraphics[width=0.8\linewidth,clip]{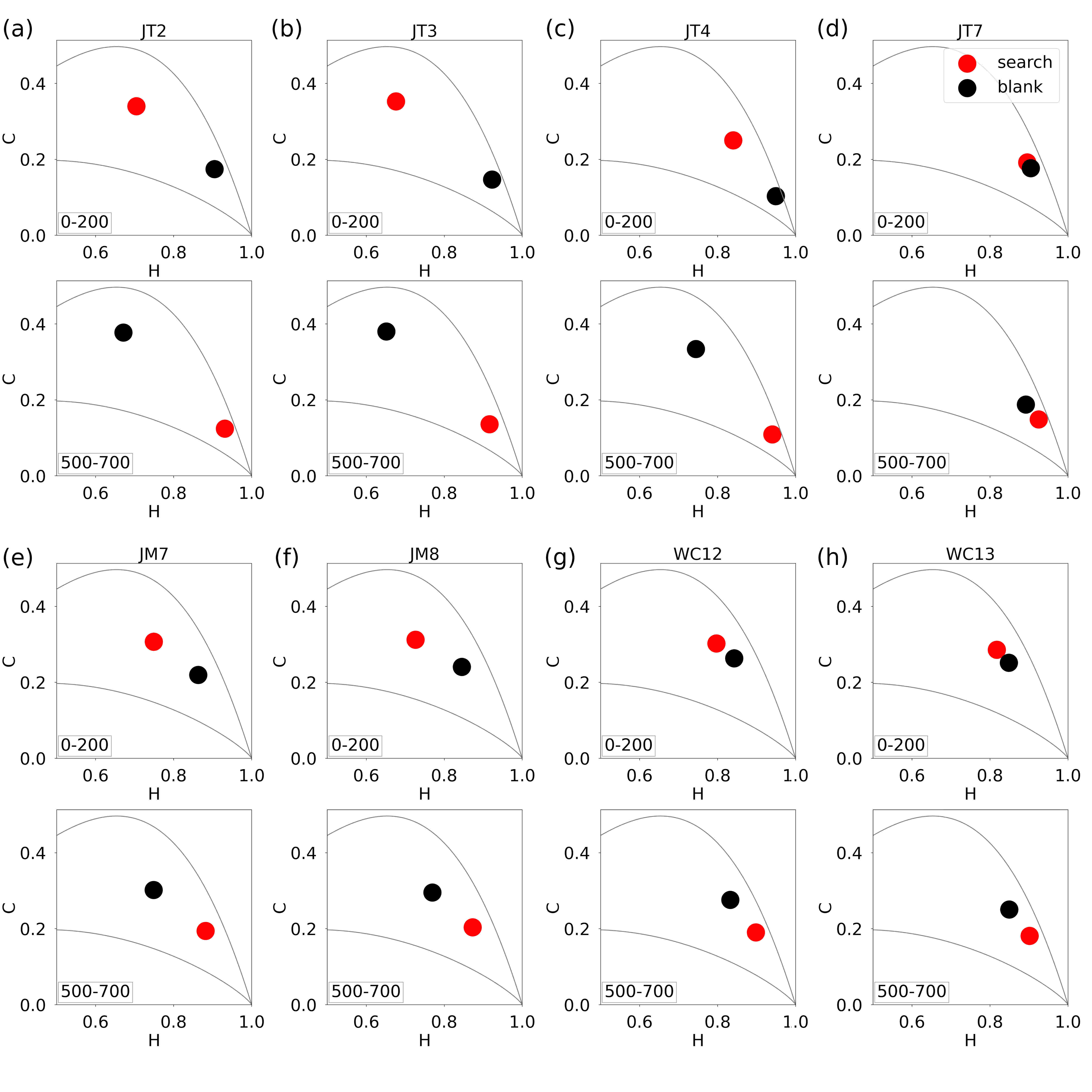}
\caption{\label{fig:HC8channels} 
Characterizing complexity and entropy for both conditions at the $C\times H$ plane in two different moments of the trials: 0 to 200ms and 500ms to 700ms. (a)-(h) The complexity-entropy plane for the 8 illustrative channels shown in Fig.~\ref{fig:Heatmap8channels}.
}
\end{figure*}

%%%%%%%%%%%%%%%%%%%%%%%%%%%%%%%%%%%%%%%%%%%%%%%%%
\begin{figure}[h!]%
 \begin{minipage}{8cm}
  \begin{flushleft}%
\end{flushleft}%
\includegraphics[width=0.99\columnwidth,clip]{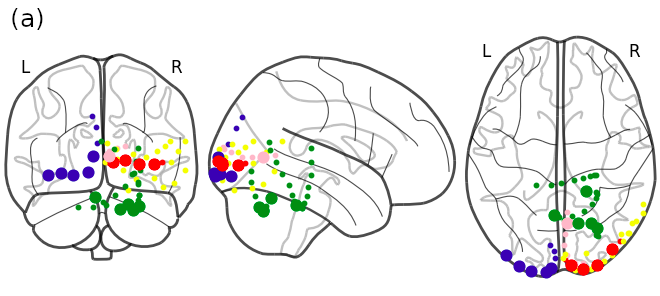}
\end{minipage}
 \begin{minipage}{8cm}
  \begin{flushleft}%
\end{flushleft}%
\includegraphics[width=0.98\columnwidth,clip]{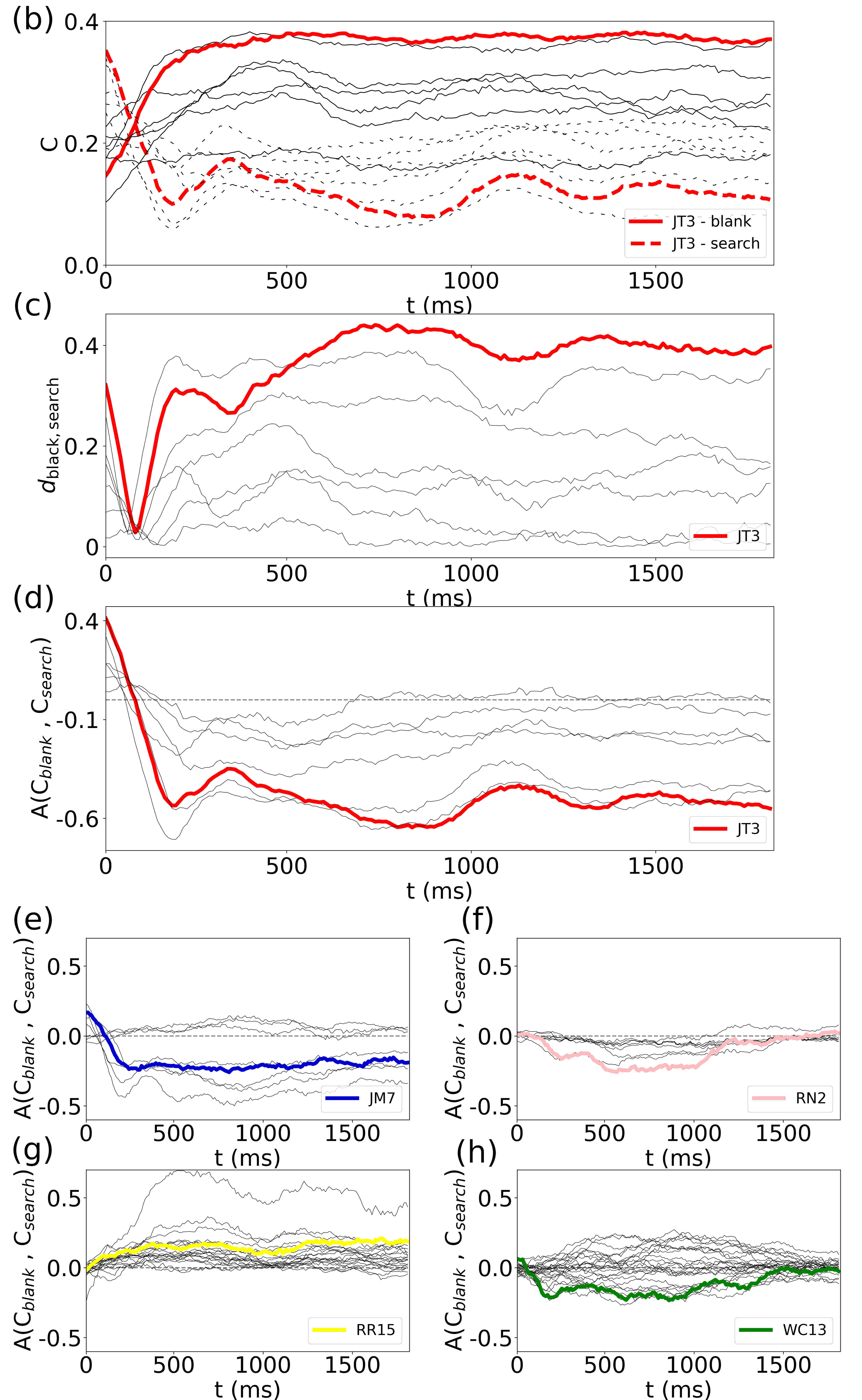}
\end{minipage}
\caption{\label{fig:CDAwindows} 
Characterizing different types of channel activit~msy for $\tau=6$~ms. 
(a) Localization of all 67 channels from five subjects (separated by the collors) in a standard brain map. Larger dots indicate similar behavior as channel JT3.
(b) Complexity $C_{search}$ and $C_{blank}$ and
(c) distance $d_{blank,search}$ for all channels from patient JT.
(d)-(h) Assimetry of complexity $A(C_{search},C_{blank})$ for all sites of each patient. The channels can be separated into three groups: positive values of asymmetry, negative values, or close to zero values during the most part of the trial. One illustrative electrodes for each patient is highlighted.
}
\end{figure}

In order to identify at a glance both the electrode channels and time scales where we can better distinguish the different trial types, in Fig. \ref{fig:heatmapallchannels} we plot the heatmap of the distance $d_{blank,search}$ for all analyzed electrodes and values of $\tau$. The vertical axis indicates the embedding time $\tau$ and the horizontal axis shows the 67 channels from five patients in the following order: 8 electrodes from subject JM, 7 from JT, 8 from RN, 20 from RR, and 24 from WC. The electrode $11$ is the exemplar JT3 electrode shown in Figures~\ref{fig:trials} and \ref{fig:HCjt3}. The color code displayed in the figure shows the magnitude of the distance $d_{blank,search}$ between trials.

As a matter of fact, in Fig. \ref{fig:heatmapallchannels} we notice that red spots on the map indicate where $d_{blank,search}$ is larger. This means that, considering the entire 2~s time series of activity at once in our analysis, only some good channels can clearly separate the two trial types. 
Moreover, the time scale around $\tau=6$~ms is a useful one to distinguish the activity type in several of these good electrodes, not only in the exemplar one. 
These pronounced differences between the two conditions for many electrodes from different patients at $\tau=6$~ms suggests that relevant information processes are happening at this time scale.

In addition, some channels from the same subject that are spatially located close to each other present considerable smoothness in the heatmap patterns, which reveals  robustness in our results.
For example, many occipital channels from JM and JT present large values of $d_{blank,search}$.
All patients present some channels that are better in distinguishing between the two conditions than others, which indicates that not all sites are engaged in the tasks.

Therefore, our results suggest that regions presenting larger distances between trial types are more involved in one of the two conditions. On one hand, sites related to the visual search task, for example, could be involved in the recognition of the arrows, the star, or the colors. On the other hand, regions related to expectation, and preparation would be more engaged in the waiting periods. Moreover, the time scales that maximize the distance would be related to specific information processes occurring in these regions.

%%%%%%%%%%%%%%%%%%%%%%%%%%%%%%%%%%%%%%%%%%

\begin{figure*}[t]%
\includegraphics[width=0.99\linewidth,clip]{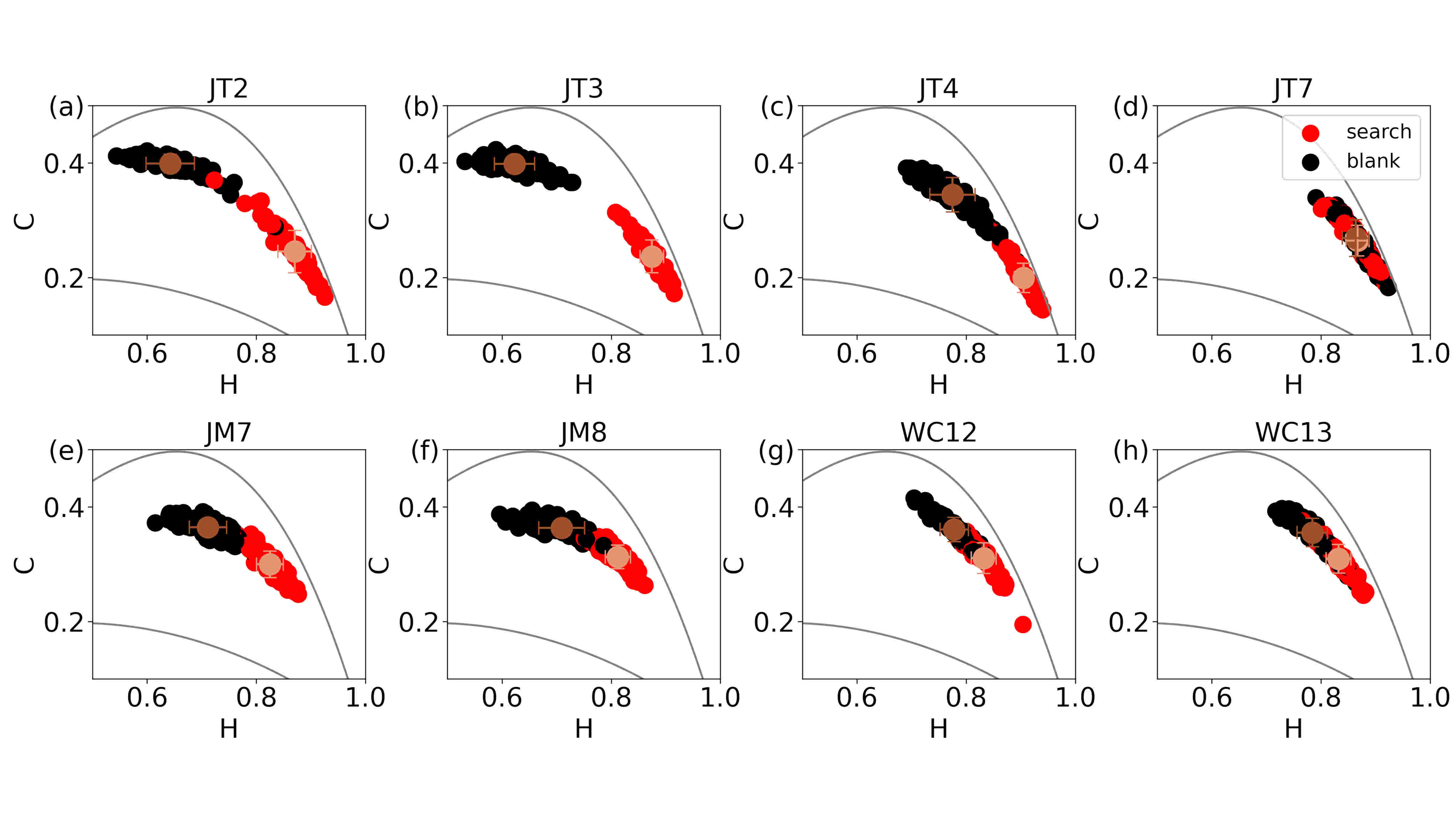}
\caption{\label{fig:HCtrialbytrial} 
Distinguishing two cognitive tasks on a trial-by-trial basis (for the same 8 illustrative channels as in Fig.~\ref{fig:Heatmap8channels}). Each black (red) dot represents entropy $H_{blank}$ ($H_{search}$) and complexity $C_{blank}$ ($C_{search}$) calculated using the 2000 points of only one trial, and $\tau=6$~ms. The brown (orange) circles and error bars are respectively the average and standard deviation in all 120 \textit{blank} (\textit{search}) trials. The exemplar electrode JT3 at the occipital area can remarkably distinguish the two conditions for all trials. 
}
\end{figure*}

%%%%%%%%%%%%%%%%%%%%%%%%%%%%%%%%%%
\subsection{Analyzing different time intervals along the 2-second-long trial duration}

One could argue that
even though the external visual stimuli remain fixed throughout the entire trial, using the 2-second window is too long considering cognitive processes. If this is the case, we are probably mixing different processes in figures \ref{fig:HCjt3} and \ref{fig:heatmapallchannels}. In particular, 
along the \textit{search} trials, many different internal processes happen in sequence: recognize the arrow direction, search for the star, and then identify the color of the correct box.
Therefore, in the following steps, we separate the 2~s time interval of each trial in intervals of $200$~ms using a sliding window of $10$~ms. Then for each interval, we repeat the analysis by calculating entropy and complexity for both conditions (\textit{blank} and \textit{search}) as a function of time. Afterward, we calculate the distance $d_{blank,search}$ and asymmetry index of the complexity $A(C_{search},C_{blank})$ between the two trial types (see equations~\ref{eq:distance} and ~\ref{eq:asymmetric-C}).

In fact, we verify for many electrodes that there are specific time intervals in which the distance between the two conditions increases.
In Fig. ~\ref{fig:Heatmap8channels} we show the heatmap of the distance $d_{blank,search}$ for $\tau$ in the y-axis and the time course along the 2~s duration of the trial in the x-axis for 8 illustrative channels. We show the exemplar electrode JT3 and its two closest neighbors (JT2 and JT4), as well as the worst electrode of this patient for distinguishing the two conditions (JT7). We also show two illustrative electrodes from patients JM and WC with similar behavior of JT3: JM7, JM8, WC12, and WC13. 
We can see that the distance between the two conditions along the time also depends on the embedding time $\tau$.
For example, many channels present a peak in the distance between $500$~ms and $1000$~ms after the beginning of the trial and for $3<\tau<12$.
The important time scales to differentiate the two conditions remain the same when we compare the results for the entire trial (Fig.~\ref{fig:heatmapallchannels}) and when we divide it into smaller windows (Fig.~\ref{fig:Heatmap8channels}). Moreover, sites that exhibit large distances only in one part of the trial have shown small values of distance in the first analysis considering the whole 2~s. This could be related to sites that are more engaged only in a very specific moment of one of the tasks, for example, the recognition of the arrow.

However, due to the design of the experiment, it is not possible to ensure that the same cognitive process happens at the same time window in every trial.
Therefore, we suggest that a sequential visual search task in which the arrow appears first on the screen and then the colored boxes appear would be useful to differentiate more subtle cognitive processes such as recognizing the arrow direction, or the color of the correct box.

It is worth mentioning that the WC patient also presents another time scale for $15<\tau<30$ which exhibits pronounced differences between the two conditions at the beginning of the trial (see Fig.~\ref{fig:Heatmap8channels}(h)). This also happens for other electrodes and it is worth more investigation since it suggests that different cognitive processes are happening in different time scales. However, hereafter we focus our analysis on $\tau=6$~ms, since the results are more pronounced and robust at this time scale for the available electrodes.

After identifying $\tau$ and time windows that are better to distinguish the two conditions,  we show in Fig. \ref{fig:HC8channels} the complexity-entropy plane for the same 8 illustrative channels of Fig. ~\ref{fig:Heatmap8channels}. We can visualize $H$ and $C$ for both conditions, $\tau=6$, and two different time intervals: 0 to 200~ms and 500~ms to 700~ms.

We verify that the analysis in small time intervals is useful not only to distinguish the two tasks but also to reveal specific characteristics of cortical activity in each condition separately. 
Naively one could initially expect that the channel activity and its indexes should not change over time 
for the \textit{blank} trials. First, because the external visual stimuli remain the same, and second because the patient is just waiting.
However, the levels of attention, expectation, and preparation could consistently change from the beginning to the end of the trial, which could influence the value of entropy and complexity.

In fact, many channels present differences in the indexes along the time intervals of the blank screen presentation around a specific time scale $\tau= 6$~ms. 
As an example, one can compare the vertical position of the black dots for the interval 0 to 200ms and 500 to 700ms in Fig.\ref{fig:HC8channels}(b) for the exemplar site JT3. 
The figure shows that $C_{blank}$ is close to 0.2 in the first analyzed interval from 0 to 200ms and close to 0.4 in the second interval from 500 to 700ms. This means that the complexity not only changes but increases during the waiting window.

We also show that at some sites, there is an inversion of the position of the \textit{blank} and \textit{search} dots in the $C\times H$ plane when comparing the two time intervals (see Fig. \ref{fig:HC8channels}). At the beginning of the trial, the position of the \textit{blank} trials in the $C\times H$ plane is closer to noisy time series. In contrast, in the second interval, the \textit{search} condition is closer to the lower right corner indicating more similarity with random data. To illustrate this inversion, compare the position of black and red dots in the interval 0 to 200ms and 500 to 700ms in Fig.\ref{fig:HC8channels}(b).

The inversion of the position of both conditions in the $C\times H$ plane, as well as the increase of $C_{blank}$ along the trial are not trivial results at all, especially because they do not happen for every value of $\tau$ and neither for every channel.
In order to better identify these changes along the trial, besides complexity and distance, we also calculate the asymmetry index for the complexity $A(C_{search},C_{blank})$ for all sites. A positive value of this index indicates $C_{search}>C_{blank}$, whereas a negative value occurs for $C_{blank}>C_{search}$. 
Therefore, the inversion of the position in the $C\times H$ plane is characterized by a change in the sign of the asymmetry index.

In Fig.~\ref{fig:CDAwindows}(a) we illustrate the position of all 67 channels in a standard brain map. The colors represent electrodes from different patients. Larger circles indicate sites that present the inversion in the $C\times H$ plane along the trial for $\tau=6$ ms.
 It is worth mentioning that many sites at occipital regions, which are related to visual processes, present similar characteristics. For example, the electrode positions of JM and JT are in equivalent occipital regions but in different hemispheres.

Therefore, in Fig.~\ref{fig:CDAwindows}(b) 
we show the complexity as a function of the time course of the 2-second trial duration for both conditions, $\tau=6$~ms, and all electrodes from patient JT. For the exemplar site JT3 (highlighted with thicker lines), $C_{blank}$ is smaller at the beginning of the trial and increases during the first hundreds of milliseconds of the trial. Then it remains reasonably stable in high values close to 0.4. 
The complexity $C_{search}$ presents an opposite behavior. It exhibits large values at the beginning of the trial and decreases along the first hundred milliseconds.  
The inversion occurs when $C_{search}=C_{blank}$ which happens between the 200~ms time window starting at 80~ms and the next one starting at 90~ms.

In Fig.~\ref{fig:CDAwindows}(c) we show 
the distance $d_{blank,search}$ along the 2-second trial duration. We can observe that for JT3 and other electrodes, the distance decreases, and goes to zero when the indexes from both conditions change position in the $C \times H$ plane. Then, the distance increases again and eventually reaches larger values than the ones at the beginning of the trial. 

To distinguish the two situations of large distance but inverted position at the plane, we plot the asymmetry index for the complexity $A(C_{search},C_{blank})$ in Fig.~\ref{fig:CDAwindows}(d). We show that the asymmetry index for JT3 starts positive and goes to negative values. 
Regarding the sign of the asymmetry index along all the time windows, the channels can be categorized into three groups: positive values, negative values, or close to zero values in the majority part of the trial. In Fig.~\ref{fig:CDAwindows}(d)-(h) we show all sites of each patient highlighting an illustrative one. We point out that only one patient (RR) does not present any electrode with consistent negative asymmetry. Moreover, 
patient WC is the only one with channels clearly representing the three conditions for the asymmetry index.

In this scenario, we hypothesize that each one of the three behaviors of channels regarding the asymmetry index present different functional roles related to the tasks. It is worth noting that, by definition, sites with $A(C_{search},C_{blank})\approx 0$ also present small values of $d_{blank,search}$. Therefore, we propose that these channels are not involved in the tasks, at least at this time scale. Moreover, 
channels with $A(C_{search},C_{blank})<0$ would be processing information more related to expectation and preparation. On the other hand, regions with $A(C_{search},C_{blank})>0$ would be more engaged in the \textit{search} task, for example, in recognizing the arrow, the star, or the colors.

%%%%%%%%%%%%%%%%%%%%%%%%%%%%%%%%%%%%%%%%%%%%%%%%%%%%%%%%%%%%%
\begin{figure}[h]%
% \begin{minipage}%{8cm}
  \begin{flushleft}%
\end{flushleft}%
\includegraphics[width=0.99\columnwidth,clip]{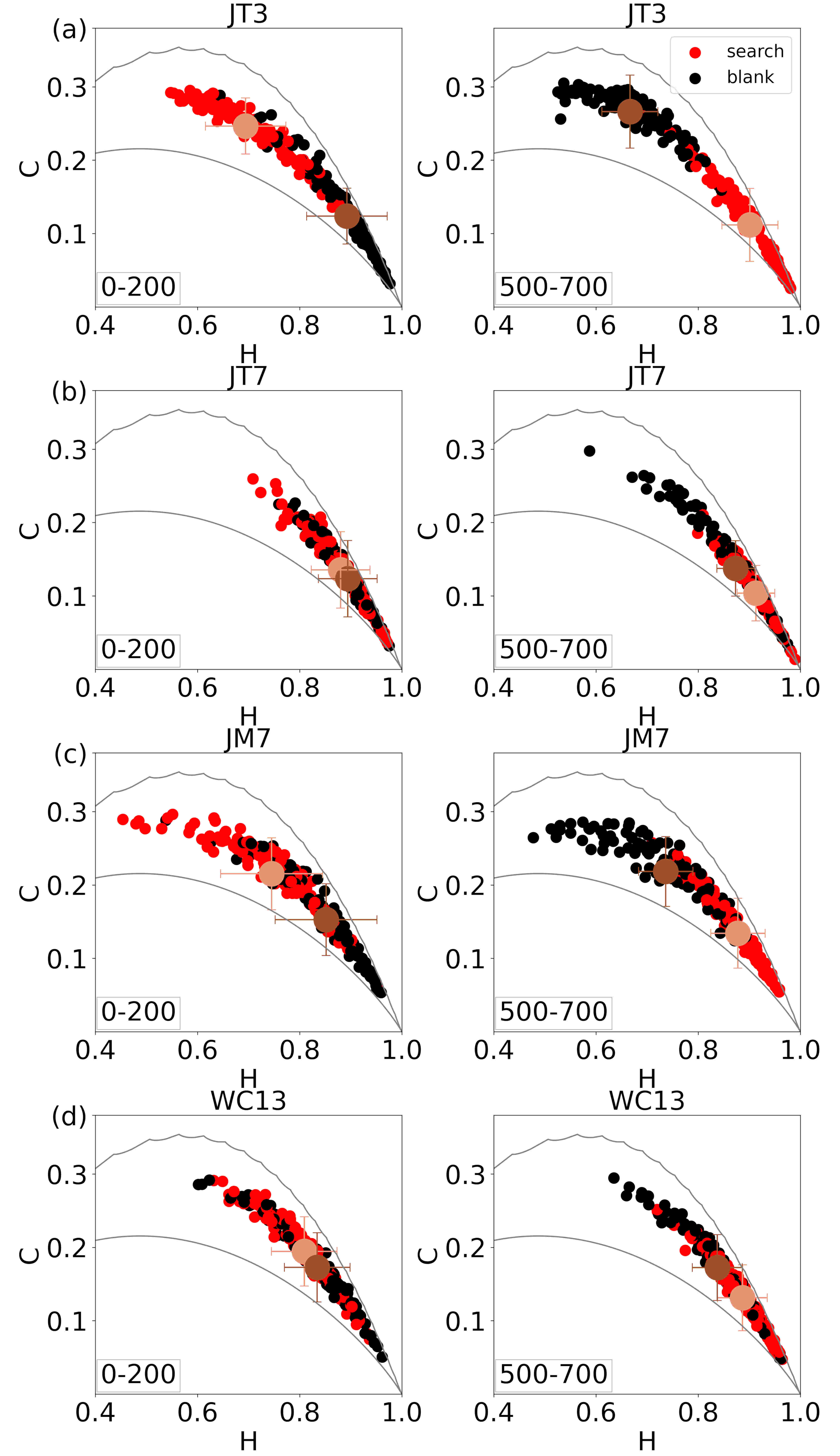}
%\end{minipage}
\caption{\label{fig:HCtrialbytrial200ms} 
Distinguishing two cognitive tasks on a trial-by-trial basis in small time windows of only 200~ms, $\tau=6$~ms, and 4 illustrative electrodes. Each black (red) dot represents entropy $H_{blank}$ ($H_{search}$) and complexity $C_{blank}$ ($C_{search}$) calculated using the time interval from 0 to 200~ms in the first column and time interval from 500 to 700~ms in the second column. The brown (orange) circles and error bars are respectively the average and standard deviation in all 120 \textit{blank} (\textit{search}) trials. 
}
\end{figure}
%%%%%%%%%%%%%%%%%%%%%%%%%%%%%%%%%%%%%%%%%%%%%%%%%%%%%%%%%%%%%

\subsection{Distinguishing cortical states on a trial-by-trial basis}
\label{subsec:trialbytrial}

In this section, we show that our method can also be applied to shorter time series, for example, on a trial-by-trial basis. For such a task, we calculate entropy and complexity for each trial independently. First, we use the whole trial, analyzing the 2000 points of each 2-second long trial. In Fig.~\ref{fig:HCtrialbytrial} we show the complexity-entropy plane for $\tau=6$~ms and the 8 illustrative electrodes shown in figures~\ref{fig:Heatmap8channels} and ~\ref{fig:HC8channels}. Each black dot represents one single \textit{blank} trial, whereas each red dot represents one \textit{search} trial. Therefore, we show 120 black dots and 120 red dots in each plot. The larger circles and error bars indicate the average values of $H$ and $C$ of the 120 dots and their standard deviations for each condition: \textit{blank} or \textit{search} trial.

In particular, for the exemplar electrode JT3, the two conditions (\textit{blank} or \textit{search} trials) are remarkably well separated in the complexity-entropy plane for all trials (see Fig.~\ref{fig:HCtrialbytrial}(b)). This means that, for this site, we would correctly guess the trial type in every attempt, by evaluating $H$ and $C$. 
Other electrodes also present good separation among trial types on average (see some examples in Fig.~\ref{fig:HCtrialbytrial} for JT2, JT4, JM7, and JM8). This means that we could guess the trial type  with a high success rate. 

As expected, the separation between the two conditions (\textit{blank} or \textit{search} tasks) does not happen for all electrodes. For example, for the JT7 channel (which has the smallest distances in Fig.~\ref{fig:Heatmap8channels}), the two conditions are totally mixed on a trial-by-trial basis. The black and red dots representing each trial from each type, as well as the averages and standard deviations for each condition, are clearly overlapping at the $C \times H$ plane. (see Fig.~\ref{fig:HCtrialbytrial}(d)).

We point out that we have checked the robustness of the results for different lengths of the ordinal patterns ($D=6,5,4,3$) to make sure that fluctuations in the probability distribution function for different numbers of possible symbols are not compromising the results. Since it is still possible to verify a clear separation between trial types for smaller values of $D$, we can also apply the method for smaller time intervals.

Considering small time windows of 200~ms of each trial independently, we can see the evolution of the separation between the two conditions along the 2-s long interval on a trial-by-trial basis. 
In Fig.~\ref{fig:HCtrialbytrial200ms} we show the $C \times H$ plane for 4 electrodes: : JT3, JT7, JM7, and WC13 (that are also shown in figures~\ref{fig:Heatmap8channels}, ~\ref{fig:HC8channels} and ~\ref{fig:HCtrialbytrial}) for $\tau=6$~ms and $D=4$. We show all 120 \textit{blank} (\textit{search}) trials in black (red) 
for two different time windows: the beginning of the trial from 0 to 200~ms in the first column; as well as for an advanced time window along the trial from 500~ms to 700~ms in the second column. 

In Fig.~\ref{fig:HCtrialbytrial200ms}(a) we show that for the exemplar channel JT3, it is still possible to distinguish the trial type on average using intervals smaller than a second-to-second basis. 
Interestingly, we can show the inversion of the position of the indexes for \textit{blank} and \textit{search} in the $C\times H$ plane comparing the two time windows (0 to 200~ms  and 500~ms to 700~ms). However, even for the best performance electrode and the best performance interval, we could not find a complete separation for all trials of each type.

%%%%%%%%%%%%%%%%%%%%%%%%%%%%%%%%%%%%%%%%%%%%%%%%

\section{Conclusion}
\label{Sec:Conclusions}
To summarize, we have shown that information theory quantifiers (such as Shannon entropy, MPR-statistical complexity, and multi-scale complexity-entropy plane~\cite{Martin06,Rosso07}) are a useful tool to characterize the information process in human intracranial signals~\cite{parvizi2018promises}. Using ECoG data from an open database~\cite{miller2010dynamic}, we show that we are able to distinguish waiting periods of the blank screen from the visual search task and infer relevant time scales for these processes.
We also determine cortical regions that present more pronounced differences between the two cognitive tasks.
Furthermore, we characterize trial-type-specific processes along different time intervals during the tasks. Finally, we show that for exemplar electrodes this can be done on a trial-by-trial basis.

Our results open new venues in the investigation of response-specific and stimulus-specific brain activity. The utilized method is potentially useful to quantify other features of the visual task such as identifying the direction of the arrow or correlating the indexes with behavioral aspects such as correct response rate. 
Differently from other classifiers, for example those which employ machine learning techniques, our method can also give us intuition about the physical significance of the signals since we can compare them to ordered and disordered states.
Moreover, we show that differences in the indexes along the time course indicate that the statistical properties of the signals are not only related to the external visual stimulus but also to the cognitive process involved in the tasks. 

We suggest that a sequential visual search task in which the arrow appears first and then the colored boxes appear would be useful to differentiate more subtle cognitive processes such as recognizing the arrow, searching for the star, and then identifying the color of the correct box. For instance, our method could be applied to understand math tasks in which the numbers appear sequentially on the screen as reported in Ref.~\cite{pinheiro2023direct,pinheiro2024spatiotemporal}.
This method can be also potentially useful to differentiate other features such as the direction of the arrow or for different tasks to distinguish numbers from letters or other symbols.

Finally, we beleive that our current analysis may also be applied to other cognitive tasks using noninvasive data such as EEG~\cite{carlos2020anticipated} and MEG~\cite{Kosem16,michalareas2016alpha}. It is also possible to employ the complexity-entropy plane to characterize differences among groups with neurodisorders~\cite{echegoyen2020permutation} and control in other to use it for helping in diagnostic or to characterize levels of anesthesia. We expect that, since even more data is becoming publicly available, it will be easier to address more sophisticated neurocognitive questions in the light of information theory methods.

%\appendix
%\section{\label{Appendix}Appendix}

\begin{acknowledgments}
The authors thank 
CNPq (grants 402359/2022-4, 314092/2021-8), FAPEAL (grant APQ2022021000015), UFAL, CAPES and L’ORÉAL-UNESCO-ABC For Women In Science (Para Mulheres na Ciência) for financial support.
\end{acknowledgments}
% 
% \pagebreak
%\appendix
\bibliography{matias}

\end{document}